\documentclass[fleqn,usenatbib]{mnras}

\usepackage{newtxtext,newtxmath}
\usepackage[T1]{fontenc}
\usepackage{ae,aecompl}
\usepackage{placeins}
\usepackage{graphicx}	 
\usepackage{amsmath}	
\usepackage{color}
\usepackage{gensymb}

\usepackage{scalerel,tikz}
\usetikzlibrary{svg.path}
\definecolor{orcidlogocol}{HTML}{A6CE39}
\tikzset{orcidlogo/.pic={
 \fill[orcidlogocol] svg{M256,128c0,70.7-57.3,128-128,128C57.3,256,0,198.7,0,128C0,57.3,57.3,0,128,0C198.7,0,256,57.3,256,128z};
 \fill[white] svg{M86.3,186.2H70.9V79.1h15.4v48.4V186.2z}
 svg{M108.9,79.1h41.6c39.6,0,57,28.3,57,53.6c0,27.5-21.5,53.6-56.8,53.6h-41.8V79.1z M124.3,172.4h24.5c34.9,0,42.9-26.5,42.9-39.7c0-21.5-13.7-39.7-43.7-39.7h-23.7V172.4z}
 svg{M88.7,56.8c0,5.5-4.5,10.1-10.1,10.1c-5.6,0-10.1-4.6-10.1-10.1c0-5.6,4.5-10.1,10.1-10.1C84.2,46.7,88.7,51.3,88.7,56.8z};
}}
\newcommand\orcidicon[1]{\href{https://orcid.org/#1}{\mbox{\scalerel*{
\begin{tikzpicture}[yscale=-1,transform shape]
\pic{orcidlogo};
\end{tikzpicture}
}{|}}}}

\newcommand{\aref}[1]{\hyperref[#1]{Appendix~\ref{#1}}}

\usepackage[normalem]{ulem}

\definecolor{darkgreen}{rgb}{0.13, 0.55, 0.13}
\definecolor{brown}{rgb}{0.65, 0.16, 0.16}

\title[Gas phase metallicity relations]{On the origin of the mass-metallicity gradient relation in the local Universe}

\author[P. Sharda et al.]{Piyush Sharda$^{\orcidicon{0000-0003-3347-7094}\,1,2}$\thanks{piyush.sharda@anu.edu.au (PS)},
Mark R. Krumholz$^{\orcidicon{0000-0003-3893-854X}\,1,2}$\thanks{mark.krumholz@anu.edu.au (MRK)},
Emily Wisnioski$^{\orcidicon{0000-0003-1657-7878}\,1,2}$\thanks{emily.wisnioski@anu.edu.au (EW)},
Ayan Acharyya$^{\orcidicon{0000-0003-4804-7142}\,1,2,3}$, 
\newauthor
Christoph Federrath$^{\orcidicon{0000-0002-0706-2306}\,1,2}$, and
John C. Forbes$^{\orcidicon{0000-0002-1975-4449}\,4}$\\
$^{1}$Research School of Astronomy and Astrophysics, Australian National University, Canberra, ACT 2611, Australia\\
$^{2}$Australian Research Council Centre of Excellence for All Sky Astrophysics in 3 Dimensions (ASTRO 3D), Australia\\
$^{3}$Department of Physics and Astronomy, Johns Hopkins University, Baltimore, MD 21218, USA\\
$^{4}$Center for Computational Astrophysics, Flatiron Institute, New York, NY 10010, USA
}

\date{Accepted 2021 March 22. Received 2021 February 18; in original form 2020 December 03}

\pubyear{2021}

\begin{document}
\label{firstpage}
\pagerange{\pageref{firstpage}--\pageref{lastpage}}
\maketitle

\begin{abstract}
In addition to the well-known gas phase mass-metallicity relation (MZR), recent spatially-resolved observations have shown that local galaxies also obey a mass--metallicity gradient relation (MZGR) whereby metallicity gradients can vary systematically with galaxy mass. In this work, we use our recently-developed analytic model for metallicity distributions in galactic discs, which includes a wide range of physical processes -- radial advection, metal diffusion, cosmological accretion, and metal-enriched outflows -- to simultaneously analyse the MZR and MZGR. We show that the same physical principles govern the shape of both: centrally-peaked metal production favours steeper gradients, and this steepening is diluted by the addition of metal-poor gas, which is supplied by inward advection for low-mass galaxies and by cosmological accretion for massive galaxies. The MZR and the MZGR both bend at galaxy stellar mass $\sim 10^{10} - 10^{10.5}\,\rm{M_{\odot}}$, and we show that this feature corresponds to the transition of galaxies from the advection-dominated to the accretion-dominated regime. We also find that both the MZR and MZGR strongly suggest that low-mass galaxies preferentially lose metals entrained in their galactic winds. While this metal-enrichment of the galactic outflows is crucial for reproducing both the MZR and the MZGR at the low-mass end, we show that the flattening of gradients in massive galaxies is expected regardless of the nature of their winds. 
\end{abstract}




\section{Introduction}
Metals have a profound impact on galaxy formation and evolution even though their contribution to the total visible matter is less than two per cent. The symbiotic relationship between galaxies and their metal content has now been investigated in detail through numerous observations, simulations and analytic models. One of the key manifestations of this relationship is the correlation between the stellar mass of a galaxy ($M_{\star}$, used as a proxy for the total galaxy mass) and its global (gas phase or stellar) metallicity, $Z$. It is now well established that low-mass galaxies have lower $Z$ as compared to massive galaxies. This is known as the mass metallicity relation \citep[MZR; e.g.,][]{2002ApJS..142...35K,2004ApJ...613..898T,2010MNRAS.408.2115M,2015Natur.521..192P,2017MNRAS.469..151B,2017ApJ...847...18Z,2017MNRAS.465.1384C,2020MNRAS.491..944C}. The exact cause of the MZR is still debated; for example, star formation \citep{2007ApJ...655L..17B}, outflows \citep{2008MNRAS.385.2181F,2018MNRAS.481.1690C}, cosmic accretion or infall \citep{1972NPhS..236....7L,2012MNRAS.421...98D}, feedback \citep{2019MNRAS.482.2208T}, and the initial mass function (IMF, \citealt{2007MNRAS.375..673K}) can all play a role in setting its shape. The shape of the MZR seen in observations has now been successfully reproduced by many simulations (e.g., \citealt{2007ApJ...655L..17B,2011MNRAS.416.1354D,2017MNRAS.467..115D,2019MNRAS.484.5587T,2019MNRAS.482.2208T}) and theoretical models (e.g., \citealt{2008MNRAS.385.2181F,2011MNRAS.417.2962P,2013ApJ...772..119L,2020MNRAS.498.3215D}); however, the absolute normalisation of the MZR (\textit{i.e.,} the absolute value of $Z$) remains uncertain due to difficulties in calibrating $Z$ from observations (see reviews by \citealt{2019ARA&A..57..511K}, \citealt{2019A&ARv..27....3M} and \citealt{2020ARA&A..58...99S}).

Since the pioneering works by \cite{1971ApJ...168..327S}, \cite{1976A&A....48..301M} and \cite{1983MNRAS.204...53S}, it has been known that galaxies also exhibit a gradient in the spatial distribution of metallicity, both in stars and in the gas phase, in the radial direction (e.g., \citealt{1994ApJ...420...87Z,2015A&A...581A.103G,2017MNRAS.466.4731G,2017MNRAS.469..151B}) as well as variations in the azimuthal direction (e.g., \citealt{2011AJ....142...51L,2013ApJ...766...17L,2017ApJ...846...39H,2019ApJ...885L..31H,2019ApJ...887...80K}). The fact that radial gradients are usually negative (\textit{i.e.,} the centre of the galaxy is more metal-rich than the outskirts) is a key piece of evidence for the theory of inside-out galaxy formation \citep{1998MNRAS.295..319M,2010PhR...495...33B,2017ARA&A..55...59N}. Hereafter, we only focus on the metallicities and metallicity gradients in the ionised gas.

Thanks to the plethora of galaxies observed in the nearby Universe with large integral field spectroscopy (IFS) surveys like CALIFA (Calar Alto Legacy Integral Field Area, \citealt{2012A&A...538A...8S}), MaNGA (Mapping nearby Galaxies at Apache Point Observatory, \citealt{2015ApJ...798....7B}), and SAMI (Sydney-AAO Multi-object Integral-field spectrograph, \citealt{2015MNRAS.447.2857B}), we can now study the trends of metallicity gradients with different galaxy properties in a statistical sense. Like the MZR, of particular interest is the stellar mass--metallicity gradient relation (MZGR). The general consensus is that the metallicity gradient, when measured in absolute units of $\rm{dex\,kpc^{-1}}$, either remains independent of stellar mass up to $M_{\star} \sim 10^{10-10.5}\,\rm{M_{\odot}}$, then flattens toward zero gradient at higher stellar masses \citep{2019A&ARv..27....3M}, or shows a mild curvature around $\sim 10^{10-10.5}\,\rm{M_{\odot}}$, with flat gradients on either side \citep[e.g.,][]{2017MNRAS.469..151B}. If the gradients are instead normalised by the effective radius of galaxies ($r_{\rm{e}}$) and expressed in $\mathrm{dex}\,r^{-1}_{\rm{e}}$, some authors find that the MZGR is steepest around $M_{\star} \sim 10^{10-10.5}\,\rm{M_{\odot}}$, with flatter gradients on either side (e.g., \citealt{2017MNRAS.469..151B,2020A&A...636A..42M,2020aMNRAS.xxx..xxxP}), whereas others report a constant, characteristic $\mathrm{dex}\,r^{-1}_{\rm{e}}$ gradient for all galaxies with $M_{\star} > 10^{9.5}\,\rm{M_{\odot}}$ \citep{2012A&A...538A...8S,2014A&A...563A..49S,2016A&A...587A..70S,2018A&A...609A.119S,2018MNRAS.479.5235P}. However, these trends in the MZGR are relatively weak as compared to the MZR, suffer observational and calibration uncertainties \citep{2013ApJ...767..106Y,2020MNRAS.495.3819A,2020MNRAS.xxx..xxxA,2020aMNRAS.xxx..xxxP}, and to date, have received limited theoretical investigation.

The goal of this work is to provide a physical explanation for the shape of the MZGR. For this purpose, we use our recently-developed first principles model of gas phase metallicity gradients \citep{2020aMNRAS.xxx..xxxS}. This model is based on the equilibrium between the production, consumption, loss and transport of metals in galactic discs. It produces gas phase metallicity gradients in good agreement with a wide range of local and high-$z$ galaxies, and shows that these gradients are in equilibrium across a diverse range of galaxy properties. We refer the reader to \cite{2020aMNRAS.xxx..xxxS} for a full description of the model, the gradients produced, as well as applications of the model to study the cosmic evolution of metallicity gradients and their trends with galaxy kinematics \citep{2020bMNRAS.xxx..xxxS}. The rest of this paper is organised as follows: \autoref{s:review} presents a review of the model, \autoref{s:MZR} describes the MZR produced by our model, which we use as a proof of concept to explain the MZGR in \autoref{s:MZGR}. \autoref{s:MZR-MZGR} introduces the MZR--MZGR space in equilibrium as a new way of characterizing gas phase metallicities, and \autoref{s:conclusions} summarizes our key results. For the purpose of this paper, we use $\rm{Z}_{\odot} = 0.0134$ for Solar metallicity, corresponding to $12 + \log_{10}\rm{O/H} = 8.69$ \citep{2009ARA&A..47..481A}, Hubble time at $z=0$: $t_{\rm{H(0)}} = 13.8\,\rm{Gyr}$ \citep{2018arXiv180706209P}, and follow the flat $\Lambda$CDM cosmology: $\Omega_{\rm{m}} = 0.27$, $\Omega_{\rm{\Lambda}} = 0.73$, $h=0.71$, and $\sigma_8 = 0.81$ \citep{2003MNRAS.339..289S}.

\section{Review of the model}
\label{s:review}
In this section, we provide a brief review of the model of gas phase metallicity gradients we presented in \cite{2020aMNRAS.xxx..xxxS}; this is intended to highlight only the results of which we will make use here, and we refer readers to the original paper for full details. In that work, we showed that the evolution of gas phase metallicity is described by the Euler-Cauchy equation
\begin{equation}
    \underbrace{\mathcal{T} s_g \frac{\partial \mathcal{Z}}{\partial \tau}}_{\substack{\text{equilibrium} \\ \text{time}}} - \underbrace{\frac{\mathcal{P}}{x} \frac{\partial \mathcal{Z}}{\partial x}}_\text{advection} - \underbrace{\frac{1}{x}\frac{\partial}{\partial x}\left(x k s_g \frac{\partial \mathcal{Z}}{\partial x}\right)}_\text{diffusion}\,\,
    = \underbrace{\mathcal{S} \dot{s}_\star}_{\substack{\text{production} \\ \text{+} \\ \text{outflows}}} - \underbrace{\mathcal{Z}\mathcal{A}\dot c_{\star}}_\text{accretion}\,,
    \label{eq:main_nondimx}
\end{equation}
where $\mathcal{Z} = Z/\mathrm{Z}_{\odot}$ is the metallicity normalised to Solar, $x$ is the radius of the disc normalised to the radius $r_0$ that we take to be the inner edge of the disc (\textit{i.e.,} $x=r/r_0$ where $r$ is the galactocentric radius), $\tau$ is the time normalised to the orbital time at $r_0$, $k$ is the normalised diffusion coefficient, and $s_g$, $\dot s_{\star}$, and $\dot{c}_\star$ are the normalised gas mass, star formation rate (SFR), and cosmic accretion rate per unit area of the galactic disc, respectively. From left to right, the different terms in \autoref{eq:main_nondimx} represent the equilibration time for a given metal distribution, radial advection of metals due to inflows, diffusion of metals due to concentration gradients, production of metals through star formation and loss via galactic outflows, and cosmic accretion of metal-poor gas from the circumgalactic medium (CGM), respectively. From \autoref{eq:main_nondimx}, we see that $\mathcal{Z}$ is governed by four dimensionless ratios. These are $\mathcal{T}$ -- the ratio of orbital to diffusion timescales, $\mathcal{P}$ -- the P\'eclet number of the galaxy that describes the ratio of advection to diffusion \citep[e.g.,][]{patankar1980numerical,rapp}, the `source' term $\mathcal{S}$ -- the ratio of metal production to diffusion, and the `accretion' term $\mathcal{A}$ -- the ratio of cosmic accretion (or infall) to diffusion.

In equilibrium, the first term goes to zero, and one can find a steady-state solution to \autoref{eq:main_nondimx} for any specified profiles of $s_g$, $\dot{s}_\star$, and $\dot{c}_\star$ versus radius. We set $s_g$ and $\dot{s}_\star$ from the unified galaxy disc model of \cite{2018MNRAS.477.2716K}, and $\dot{c}_\star$ based on cosmological simulations \citep[e.g.,][]{2008A&A...483..401C}. For these choices, the corresponding equilibrium solution for the metallicity as a function of normalised galactocentric radius, $\mathcal{Z}(x)$, is given by
\begin{eqnarray}
    \lefteqn{\mathcal{Z}(x) = \frac{\mathcal{S}}{\mathcal{A}} + c_1 x^{\frac{1}{2}\left[\sqrt{\mathcal{P}^2+\,4\mathcal{A}}-\mathcal{P}\right]}
    }
    \nonumber \\
    & & {}  + \left(\mathcal{Z}_{r_0} - \frac{\mathcal{S}}{\mathcal{A}} - c_1\right) x^{\frac{1}{2}\left[-\sqrt{\mathcal{P}^2+\,4\mathcal{A}}-\mathcal{P}\right]},
    \label{eq:Zequation}
\end{eqnarray}
where $c_1$ is a constant of integration that is determined by the metallicity of the CGM, $\mathcal{Z}_{\rm CGM}$, and $\mathcal{Z}_{r_0}$ is the equilibrium metallicity at $r_0$ that we can determine from other galaxy parameters. We can also express $\mathcal{P},\,\mathcal{S}$ and $\mathcal{A}$ in terms of meaningful galaxy parameters using the \citet{2018MNRAS.477.2716K} model, which gives
\begin{equation}
    \mathcal{T} = \frac{3\phi_Q\sqrt{2\left(\beta+1\right)}f_{g,Q}}{Q_{\rm min}}\left(\frac{v_{\phi}}{\sigma_g}\right)^2\,,
    \label{eq:physicalChi}
\end{equation}
\begin{equation}
    \mathcal{P} = \frac{6\eta\phi^2_Q\phi^{3/2}_{\rm{nt}} f^2_{g,Q}}{Q^2_{\rm{min}}}\left(\frac{1+\beta}{1-\beta}\right)\left(1 - \frac{\sigma_{\mathrm{sf}}}{\sigma_g}\right)\,,
    \label{eq:physicalP}
\end{equation}
\begin{equation}
    \mathcal{S} = \frac{24 \phi_Q  f^2_{g,Q} \epsilon_{\rm{ff}} f_{\rm{sf}}}{\pi Q_{\rm min} \sqrt{3f_{g,P} \phi_{\rm{mp}}}}\left(\frac{\phi_y y}{Z_{\odot}}\right)\left(1+\beta\right)\left(\frac{v_{\phi}}{\sigma_g}\right)^2\,,
    \label{eq:physicalS}
\end{equation}
\begin{equation}
    \mathcal{A} = \frac{3G\dot M_{h}f_{\mathrm{B}}\epsilon_{\mathrm{in}}\phi_Q}{2\sigma^3_g \left[\ln x_{\mathrm{max}} - \ln x_{\mathrm{min}}\right]}\,.
    \label{eq:physicalA}
\end{equation}
Here, $\phi_Q-1$ is the ratio of the gas to stellar Toomre $Q$ parameters \citep{1994ApJ...427..759W,2011MNRAS.416.1191R,2013MNRAS.433.1389R}, $\beta$ is the rotation curve index of the galaxy, $f_{g,Q}$ and $f_{g,P}$ are two slightly different measures of the effective gas fraction \citep{2010ApJ...721..975O,2018MNRAS.477.2716K}, $Q_{\rm{min}}$ is the Toomre $Q$ parameter \citep{1964ApJ...139.1217T} below which discs are unstable due to gravity \citep[e.g.,][]{2010ApJ...724..895K,2015ApJ...814..131G}, $v_{\phi}$ is the rotational velocity of the galaxy, $\sigma_g$ is the gas velocity dispersion, $\eta$ is a dimensional factor of order unity describing the rate of turbulent dissipation \citep{1998PhRvL..80.2754M,2012ApJ...754...48F}, $\phi_{\rm{nt}}$ is the fraction of total velocity dispersion that is in non-thermal rather than thermal motions, $\sigma_{\rm{sf}}$ is the maximum velocity dispersion that can be maintained by star formation feedback, $\epsilon_{\rm{ff}}$ is the star formation efficiency per free-fall time \citep{2005ApJ...630..250K,2012ApJ...761..156F,2012ApJ...759L..27P}, $f_{\rm{sf}}$ is the fraction of gas that is molecular \citep{2009ApJ...693..216K,2013MNRAS.436.2747K}, $\phi_{\rm{mp}}$ is the ratio of the total to the turbulent pressure at the mid-plane \citep{2010ApJ...721..975O}, $\dot M_{h}$ is the dark matter accretion rate onto the halo \citep{2008MNRAS.383..615N,2010ApJ...718.1001B}, $f_{\rm{B}}$ is the universal baryonic fraction \citep{1995MNRAS.273...72W,2016A&A...594A..13P}, and $\epsilon_{\rm{in}}$ is the baryonic accretion efficiency \citep{2011MNRAS.417.2982F}. We refer the readers to \citet[Tables 1 and 2]{2020aMNRAS.xxx..xxxS} for full descriptions of and typical values for all these parameters.

In addition to these quantities, the production term $\mathcal{S}$ depends on one additional parameter: the yield reduction factor $\phi_y$, which describes the reduction in the metal yield due to preferential ejection of metals through galactic outflows. $\phi_y = 1$ corresponds to metals injected by Type II supernovae fully mixing with the interstellar medium (ISM), while $\phi_y = 0$ corresponds to all newly produced Type II supernovae metals being ejected from the galaxy immediately, without ever becoming part of the ISM.\footnote{It is important to clarify that $\phi_y$ is not the same as the metal outflow rate or the metal mass loading factor, since $\phi_y$ only describes how metals are partitioned between winds and the ISM, not the total metal mass carried by the winds. For example, a galaxy could have very low mass loading but also low $\phi_y$, if the winds consisted primarily of metal-rich supernova ejecta, with very little additional ISM mass entrained.}

The \citet{2020aMNRAS.xxx..xxxS} model is distinct from earlier models for galaxy metallicity distributions in a few ways: (1.) we include all major transport processes, including advection and diffusion of metals, both of which are usually neglected, but which can become important in some regimes, as we show below; (2.) we do not make the common assumption that the wind and ISM metallicities are equal, since there is observational evidence that they are not \citep[e.g.,][]{2002ApJ...574..663M,2009ApJ...697.2030S,2018MNRAS.481.1690C}; (3.) we derive model parameters such as the star formation rate, radial advection rate, diffusion rate, etc., from a physical model of galactic discs that is well tested against observations \citep{Johnson18a,2019MNRAS.486.4463Y,2019ApJ...880...48U,2020MNRAS.495.2265V,2021arXiv210104122G,2020bMNRAS.xxx..xxxS}, rather than adopting parameterised prescriptions of unknown accuracy; (4.) our model allows us to study both global and spatially-resolved metallicity properties. 

However, the model also has some important limitations that we should note. First, we derive solutions for $\mathcal{Z}(x)$ only for galaxies whose metal distributions are in equilibrium; we show in \cite{2020aMNRAS.xxx..xxxS} that almost all galaxies at $z=0$ except ongoing mergers satisfy this requirement, as do the majority of galaxies out to at least $z\approx 3$. However, a major exception to this \textit{may be} galaxies with inverted gradients; for this reason we do not study inverted gradients with this model. We also make a number of simplifying assumptions in order to obtain our analytic solutions: we assume that the rotation curve index $\beta$ is a constant. We use the instantaneous recycling approximation \citep{1980FCPh....5..287T}, which means that the model is best applied to elements that are returned to the ISM quickly via Type II supernovae, rather than over longer time scales by other nucleosynthetic sources. We assume gas accreting onto the galaxy can be described by a single, fixed metallicity, which implicitly means that we neglect galactic fountains, long-term wind recycling through the CGM, and other environmental effects (e.g., the presence of satellites). Nonetheless, as we show in the next three sections that the model can successfully explain the MZR (\autoref{s:MZR}), the MZGR (\autoref{s:MZGR}), and the relationship between the two (\autoref{s:MZR-MZGR}).

\begin{figure*}
\includegraphics[width=\linewidth]{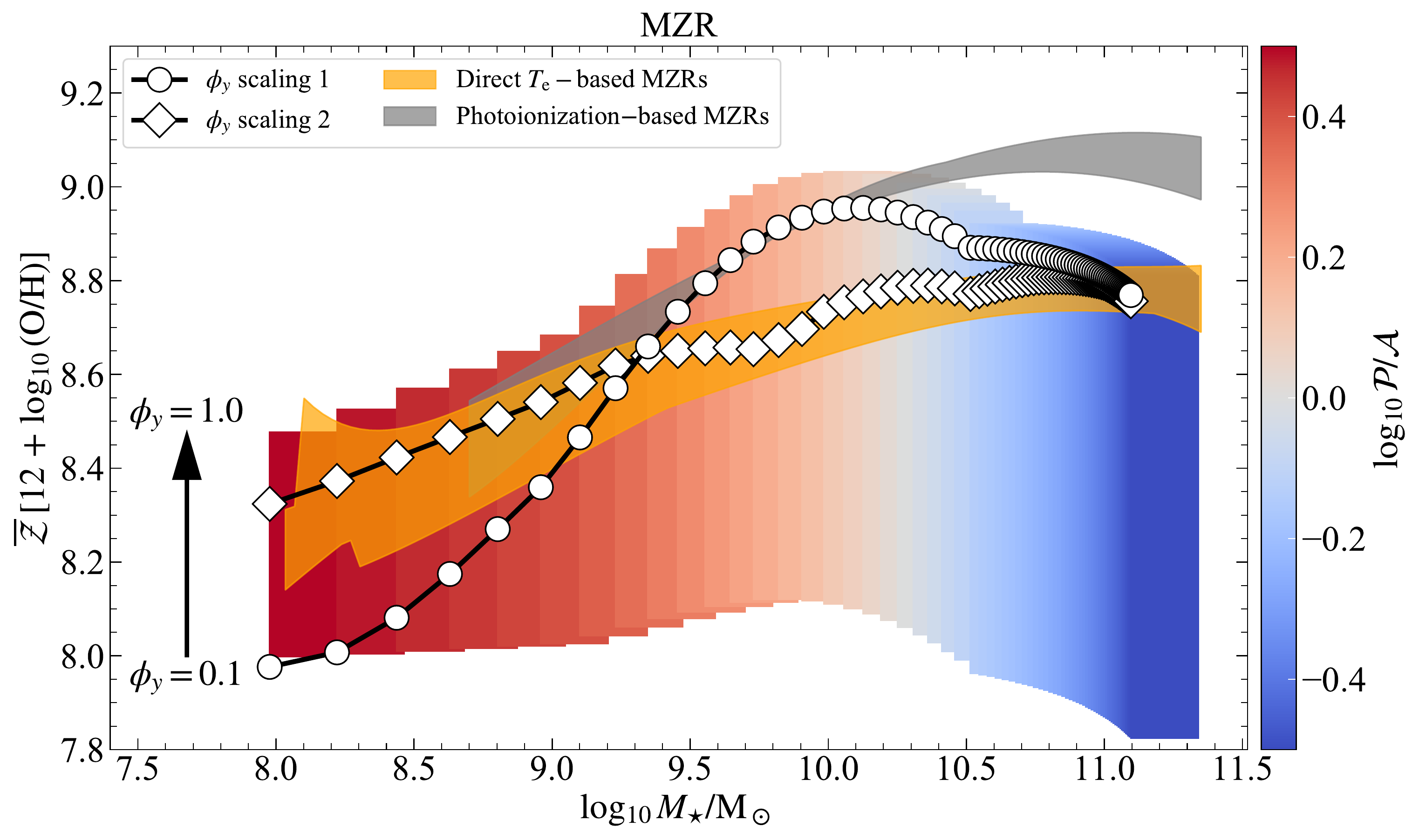}
\caption{Mass--metallicity relation (MZR) in local galaxies predicted by the \protect\cite{2020aMNRAS.xxx..xxxS} model, for different yield reduction factors $\phi_y$, color-coded by the ratio of the Péclet number ($\mathcal{P}$) to cosmic accretion over diffusion ($\mathcal{A}$). The MZR displays a curvature around $M_{\star} \sim 10^{10}-10^{10.5}\,\rm{M_{\odot}}$, corresponding to the transition from the advection-dominated ($\mathcal{P}>\mathcal{A}$) to the accretion-dominated ($\mathcal{P}<\mathcal{A}$) regime. Overlaid on the model are parameter spaces corresponding to MZRs derived from observations, using the direct $T_{\mathrm{e}}$ method \protect\citep{2004MNRAS.348L..59P,2013ApJ...765..140A,2017MNRAS.465.1384C,2020MNRAS.491..944C}, and photoionization models \protect\citep{2002ApJS..142...35K,2004ApJ...613..898T,2010MNRAS.408.2115M}, adopted from \protect\cite[Figure 15]{2019A&ARv..27....3M}. Finally, the white markers show model predictions using two possible empirical scalings of $\phi_y$ with $M_\star$. Scaling 1 is derived from observations \protect\citep{2018MNRAS.481.1690C}, whereas scaling 2 is independently derived from the best match between the model MZR and the \protect\cite{2020MNRAS.491..944C} MZR; details of these scalings are given in \aref{s:app_phiy_mstar}. Our findings predict a scaling of $\phi_y$ with $M_{\star}$ where massive galaxies prefer a higher value of $\phi_y$, and vice-versa. This implies that low-mass galaxies have more metal-enriched winds, consistent with observations \protect\citep{2018MNRAS.481.1690C} and simulations \citep{2018ApJ...869...94E,2020ApJ...899..108T}.}
\label{fig:mzr}
\end{figure*}

\section{Mass--metallicity relation (MZR)}
\label{s:MZR}

\subsection{Results on the MZR from the model}
\label{s:MZR_ours}
Almost all the analytic models that reproduce the observed MZR do not have spatial information of the distribution of metallicities in a galaxy -- these are typically developed to study global metallicities in galaxies. Although the primary focus of our work is to explain metallicity gradients by making use of the spatial information of metallicity, our model also reproduces the MZR as a proof of concept.  

To produce an MZR from the model, we need an estimate of the mean metallicity in galaxies as a function of $M_{\star}$. For this purpose, we use the SFR-weighted mean metallicity given by \citet[equation~46]{2020aMNRAS.xxx..xxxS}
\begin{equation}
\overline{\mathcal{Z}} = \frac{\int^{x_\mathrm{max}}_{x_{\mathrm{min}}} x \dot s_{\star} \mathcal{Z} dx}{\int^{x_\mathrm{max}}_{x_{\mathrm{min}}} x \dot s_{\star} dx}\,,
\label{eq:Zmean}
\end{equation}
where $\dot s_{\star}(x) = 1/x^2$ is the radial distribution of star formation per unit area \citep{2018MNRAS.477.2716K}. We use the SFR-weighted $\overline{\mathcal{Z}}$, because it can be directly compared against available MZRs since they are inherently sensitive to the SFR as the nebular metallicities are measured in H~\textsc{ii} regions around young stars \citep{2014ApJ...791..130Z}. Additionally, semi-analytic models and simulations too use SFR-weighted metallicities to construct MZRs (e.g., \citealt{2019MNRAS.482.2208T,2019MNRAS.484.5587T,2019MNRAS.487.3581F,2020arXiv201104670Y}). 

In order to derive results in terms of $M_{\star}$, we treat the rotational velocity, $v_{\phi}$, as the primary quantity that we vary. For each $v_{\phi}$, we can estimate the corresponding halo mass $M_{\rm{h}}$ and halo accretion rate $\dot{M}_{\rm h}$ at $z=0$ \citep[equations 34--35]{2020aMNRAS.xxx..xxxS}. We convert the halo mass to $M_{\star}$ following the $M_{\mathrm{h}}-M_{\star}$ relation from \cite{2013MNRAS.428.3121M} for the local Universe. Following \cite{2020aMNRAS.xxx..xxxS}, we keep the yield reduction factor, $\phi_y$, as a free parameter and vary it between 0.1 and 1, though we note that, based on both theory and observations, $\phi_y$ is expected to be close to unity in massive galaxies. For all other parameters, in particular the velocity dispersion $\sigma_g$, we use the fiducial values listed in \citet[Tables~1 and~2]{2020aMNRAS.xxx..xxxS}. Specifically, we use local dwarf values for galaxies with $M_{\star} \leq 10^9\,\rm{M_{\odot}}$, and local spiral values for $M_{\star} \geq 10^{10.5}\,\rm{M_{\odot}}$. For intermediate stellar masses, we linearly interpolate in $\log_{10} M_\star$ between these two limits for all parameters. For example, the velocity dispersions we adopt for spirals and dwarfs are $10\,\rm{km\,s^{-1}}$ and $7\,\rm{km\,s^{-1}}$ respectively, so we adopt $\sigma_g = \left(2\log_{10} M_{\star}/\mathrm{M_{\odot}} - 11\right)\,\mathrm{km\,s^{-1}}$ for intermediate-mass galaxies with $10^9\,\mathrm{M_{\odot}} < M_{\star} < 10^{10.5}\,\mathrm{M_{\odot}}$. We have verified that the resulting MZR and MZGR are not particularly sensitive to the choice of the $M_{\star}$ boundaries invoked to classify dwarfs and spirals; we also discuss this further in \autoref{s:MZGR}. We set $\mathcal{Z}_{r0}$ to its equilibrium value \citep{2020aMNRAS.xxx..xxxS}, and set the circumgalactic medium metallicity to $\mathcal{Z}_{\rm{CGM}}=0.2$ for all galaxies\footnote{This is slightly lower than the median $\mathcal{Z}_{\rm{CGM}}=0.3$ found by \cite{2017ApJ...837..169P} for $z\sim0.2$ galaxies (see also, \citealt{2016ApJ...831...95W}, where the authors find a bimodal distribution of $\mathcal{Z}_{\rm{CGM}}$); however, these surveys do not cover the entire range in galaxy masses we are interested in, and we expect $\mathcal{Z}_{\rm{CGM}}$ to be lower in low mass galaxies. In any case, this difference does not have a significant effect on the MZGR.}, which sets $c_1$. The MZR (as well as the MZGR discussed below) is insensitive to $\mathcal{Z}_{r_0}$ and only weakly sensitive to $\mathcal{Z}_{\rm{CGM}}$ as compared to $\phi_y$, so we do not vary $\mathcal{Z}_{\rm GCM}$ separately. Finally, we follow \cite{2014ApJ...788...28V} to estimate $r_{\rm{e}}$ as a function of $M_{\star}$, and set $x_{\mathrm{min}}=0.5\,r_{\mathrm{e}}$ and $x_{\mathrm{max}}=3\,r_{\mathrm{e}}$ as the range of radii $x$ over which our model solution applies. This range of radii roughly mimics that over which metallicities are measured.

\autoref{fig:mzr} shows the resulting MZR from our model, color-coded by the ratio $\mathcal{P}/\mathcal{A}$ that describes the relative strength of advection to cosmic accretion. We remind the reader that both $\mathcal{P}$ and $\mathcal{A}$ (as well as $\mathcal{S}$) are normalised by diffusion in the model. The vertical spread in the model MZR is a result of varying $\phi_y$. We also overplot the parameter space of observed MZRs from several other works based on the direct $T_{\rm{e}}$ method \citep{2004MNRAS.348L..59P,2013ApJ...765..140A,2017MNRAS.465.1384C,2020MNRAS.491..944C} and photoionization modeling \citep{2002ApJS..142...35K,2004ApJ...613..898T,2010MNRAS.408.2115M}, all of which we adopt from \citet[Figure~15]{2019A&ARv..27....3M}. We see that the model is able to reproduce the MZR of the local Universe albeit with a large spread due to $\phi_y$. There are several factors behind quantitative differences between the model MZR and MZRs in the literature. From the perspective of the model, these differences are attributed to the choice of the metal yield $y$, excluding the galaxy nucleus while finding the mean metallicities, and the absolute size of the galaxy disc. From the perspective of the MZRs we compare the model with, these differences are due to calibration and observational uncertainties, as well as limited coverage of the galaxy discs. 

In order to match with the measured MZRs, the model prefers higher $\phi_y$ for massive galaxies and lower $\phi_y$ for low-mass galaxies. This implies that metals are well-mixed in the ISM in massive galaxies before they are ejected through outflows, whereas in dwarf galaxies, some fraction of metals are ejected directly before they can mix in the ISM; in other words, the best match between the model MZR and the literature MZRs predicts that dwarf galaxies have more metal-enriched winds than massive galaxies. This finding is not new and has been theorized in several works \citep[e.g.,][]{1974MNRAS.169..229L,1986ApJ...303...39D,2007ApJ...658..941D,2008MNRAS.385.2181F,2013ApJ...772..119L,2013MNRAS.430.2891D,2019MNRAS.487.3581F}, simulations \citep{2015MNRAS.446.2125C,2016MNRAS.456.2140M,2018ApJ...867..142C,2018ApJ...869...94E,2019MNRAS.482.1304E}, and also has some observational evidence \citep{2002ApJ...574..663M,2018MNRAS.481.1690C}. 

To further treat the question of how $\phi_y$ scales with $M_{\star}$ quantitatively, we also plot two models for this scaling. We obtain the first of these from available observations that directly constrain the ratio of wind metallicity to ISM metallicity \citep{2018MNRAS.481.1690C}, and the second simply by forcing the model to reproduce the observed MZR provided by \cite{2020MNRAS.491..944C}. \aref{s:app_phiy_mstar} describes how we obtain these scalings (and the associated uncertainties) in detail. While the shape of the first scaling is consistent with observed MZRs, the second is almost identical to the direct $T_{e}$ based MZRs by construction; we include the second scaling nonetheless because there is no guarantee that the scaling we have enforced to produce the MZR will also yield the correct MZGR, a question we explore below.

\autoref{fig:mzr} shows that the MZR bends roughly where the ratio $\mathcal{P}/\mathcal{A}$ passes through unity. We can understand this behaviour as follows: the total metallicity is set by a competition between metal production (the term $\mathcal{S}$) and dilution by metal-poor gas, which can be supplied either by direct cosmological accretion onto the disc ($\mathcal{A}$) or advection of gas from the weakly-star-forming outskirts to the more rapidly-star-forming centre ($\mathcal{P}$). Each of these terms varies differently with rotation curve velocity $v_\phi$, which in turn correlates with stellar mass; as shown in \citet{2020aMNRAS.xxx..xxxS}, $\mathcal{P}$ is independent of $v_{\phi}$,\footnote{Recall that each of these terms is expressed as the relative importance of a particular process compared to metal diffusion; thus, $\mathcal{P}\propto v^0_{\phi}$ does not mean that advection is equally rapid in all galaxies independent of stellar mass, just that the ratio of advection to diffusion does not explicitly depend on stellar mass.} while $\mathcal{S} \propto v^2_{\phi}$ and $\mathcal{A} \propto v^{3.3}_{\phi}$. In the low-mass regime, corresponding to small $v_{\phi}$, we have $\mathcal{P}>\mathcal{A}$, implying that the metallicities are primarily set by the balance between source and advection. Since $\mathcal{P}\propto v^0_{\phi}$ and $\mathcal{S}\propto v^2_{\phi}$, as we go to smaller $M_{\star}$ and $v_{\phi}$, the equilibrium metallicity drops because of lower $v_{\phi}$ and lower $\phi_y$ as compared to massive galaxies. On the contrary, in the high-mass regime $\mathcal{A}>\mathcal{P}$, implying that the metallicities are set by the balance between $\mathcal{A}$ and $\mathcal{S}$. Since $\mathcal{A}\propto v^{3.3}_{\phi}$, which is stronger than the dependence of $\mathcal{S}$ on $v_{\phi}$, the metallicity, which is proportional to $\mathcal{S}/\mathcal{A}$, ceases to rise with $M_\star$, and instead reaches a maximum and starts to decrease. However, the decrease is rather mild, because shortly after passing the value of $v_\phi$ where we move into the $\mathcal{A}>\mathcal{P}$ regime, galaxies become so massive that they cease to be star-forming altogether. Thus, among star-forming galaxies, the trend of $\overline{\mathcal{Z}}$ with $M_\star$ is simply that $\overline{\mathcal{Z}}$ ceases to increase and reaches a plateau. For less massive galaxies the dominant source of metal-poor gas is advection rather than accretion. However, this only holds as long as advection is non-zero; for low mass galaxies where there is no advection (\textit{i.e.,} there is no turbulence due to gravity), it falls upon cosmic accretion to balance metal production. Since cosmic accretion is much weaker in low mass galaxies, it can take a long time for this balance to approach a steady-state, which can push the gradients out of equilibrium \citep[Section~5.1]{2020aMNRAS.xxx..xxxS}.

\subsection{Comparison with previous work}
\label{s:compare_mzr}
The existence of a local gas phase MZR has been known since early analysis of data from the Sloan Digital Sky Survey \citep[SDSS,][]{2004ApJ...613..898T}, although the absolute normalisation of the MZR remains an unsolved issue due to systematic calibration uncertainties \citep{2008ApJ...681.1183K,2016MNRAS.457.3678P,2016MNRAS.458.1529B,2017MNRAS.465.1384C,2017ApJ...844...80B,2021arXiv210207058T}. Despite these uncertainties, however, it is clear both that a relationship exists, and that it has a characteristic mass scale of $\sim 10^{10.5}$ M$_\odot$ at which the curvature of the relation changes \citep{2019ApJ...877....6B}. Not surprisingly, there have been numerous attempts to explain these relations theoretically, and it is interesting to put our model in the context of these works. However, we caution that what follows is only a partial discussion of the (vast) literature on this topic, and refer readers to the comprehensive review by \citet[Section~5.1]{2019A&ARv..27....3M}.

The basic result from theoretical models to date is that galaxies tend to approach equilibrium between inflows, accretion, star formation and outflows, which naturally gives rise to the observed MZR \citep{2008MNRAS.385.2181F,2012MNRAS.421...98D,2013ApJ...772..119L,2013MNRAS.430.2891D,2014MNRAS.443..168F}. Our results are broadly consistent with this picture. However, there are some subtle differences among published models, and between existing models and ours. One important point of distinction is the extent to which outflows are metal-enriched relative to the ISM (\textit{i.e.,} $\phi_y<1$ in the language our model), and whether this enrichment varies as a function of galaxy mass or other properties (as is the case for our two possible scalings). As already discussed, many authors simply assume that outflows are not metal-enriched (\textit{i.e.,} the outflow metallicity is the same as the ISM metallicity, $\phi_y = 1$ in our notation; e.g., \citealt{2008MNRAS.385.2181F,2012MNRAS.421...98D,2015MNRAS.446..521S,2016MNRAS.461.1760H,2017MNRAS.467..115D, 2018MNRAS.481..954C, 2020MNRAS.498.3215D}), and produce an MZR based on this assumption. Others explicitly contemplate values of $\phi_y<1$ \citep[e.g.,][]{2007ApJ...658..941D,2010A&A...514A..73S,2011MNRAS.417.2962P,2015ApJ...808..129L, 2014MNRAS.443..168F, 2019MNRAS.487.3581F, 2020arXiv201104670Y, 2021arXiv210204135K}. Our conclusion that reproducing the full shape of the MZR requires $\phi_y < 1$, particularly in low-mass galaxies, is consistent with the findings of the latter group of investigators. However, many of these authors do not study the relative importance of metal-enriched outflows for dwarfs versus spirals, which we find to be important.

It is also debated whether the MZR really has a curvature at intermediate stellar masses, and if it does, whether it simply flattens out or starts to bend. While some simulations do find curvature in the MZR around $10^{10}-10^{10.5}\,\rm{M_{\odot}}$ \citep[e.g.,][]{2017MNRAS.467..115D,2019MNRAS.484.5587T}, others do not \citep[e.g.,][]{2014MNRAS.438.1985T,2015MNRAS.452..486D,2016MNRAS.456.2140M}. Our model is consistent with the former, especially if we look at the empirical scalings of $\phi_y$ with $M_{\star}$. Moreover, recent results also show that the curvature is physical and persists in the data even after observational uncertainties are accounted for \citep{2019ApJ...877....6B}. However, the cause behind the curvature is not completely understood, and factors like Active Galactic Nuclei (AGN) feedback \citep{2017MNRAS.472.3354D}, gas recycling \citep{2014MNRAS.443.3809B}, effective gas fraction \citep{2019MNRAS.484.5587T}, chemical saturation in the ISM of massive galaxies \citep{2013ApJ...771L..19Z}, and a transition in galaxy regimes together with metal-enriched outflows as we show in this work can all play a role.

In addition to the models above, to which our results are directly comparable, a number of authors have studied the dependence of the MZR on factors not included in our work, like downsizing, time-dependent outflows, variations in star formation efficiencies and IMF, presence of satellites, environmental effects, etc. \citep[e.g.,][]{2007MNRAS.375..673K,2008MNRAS.390..245C,2008A&A...488..463M,2009A&A...504..373C,2010A&A...514A..73S,2010ApJ...718.1001B,2013A&A...550A.115H,2014MNRAS.438..262P,2016ApJ...822..107G,2017MNRAS.468.1881W,2017MNRAS.464..508B,2018MNRAS.474.1143L,2018MNRAS.476.3883L}. However, unlike the current work, most models only study the MZR and not the MZGR, thus it is difficult to reconcile whether their conclusions hold or are self-consistent with spatially-resolved galaxy properties.


\section{Mass-metallicity gradient relation (MZGR)}
\label{s:MZGR}

\subsection{Results on the MZGR from the model}
\label{s:MZGR_ours}
\begin{figure*}
\includegraphics[width=\linewidth]{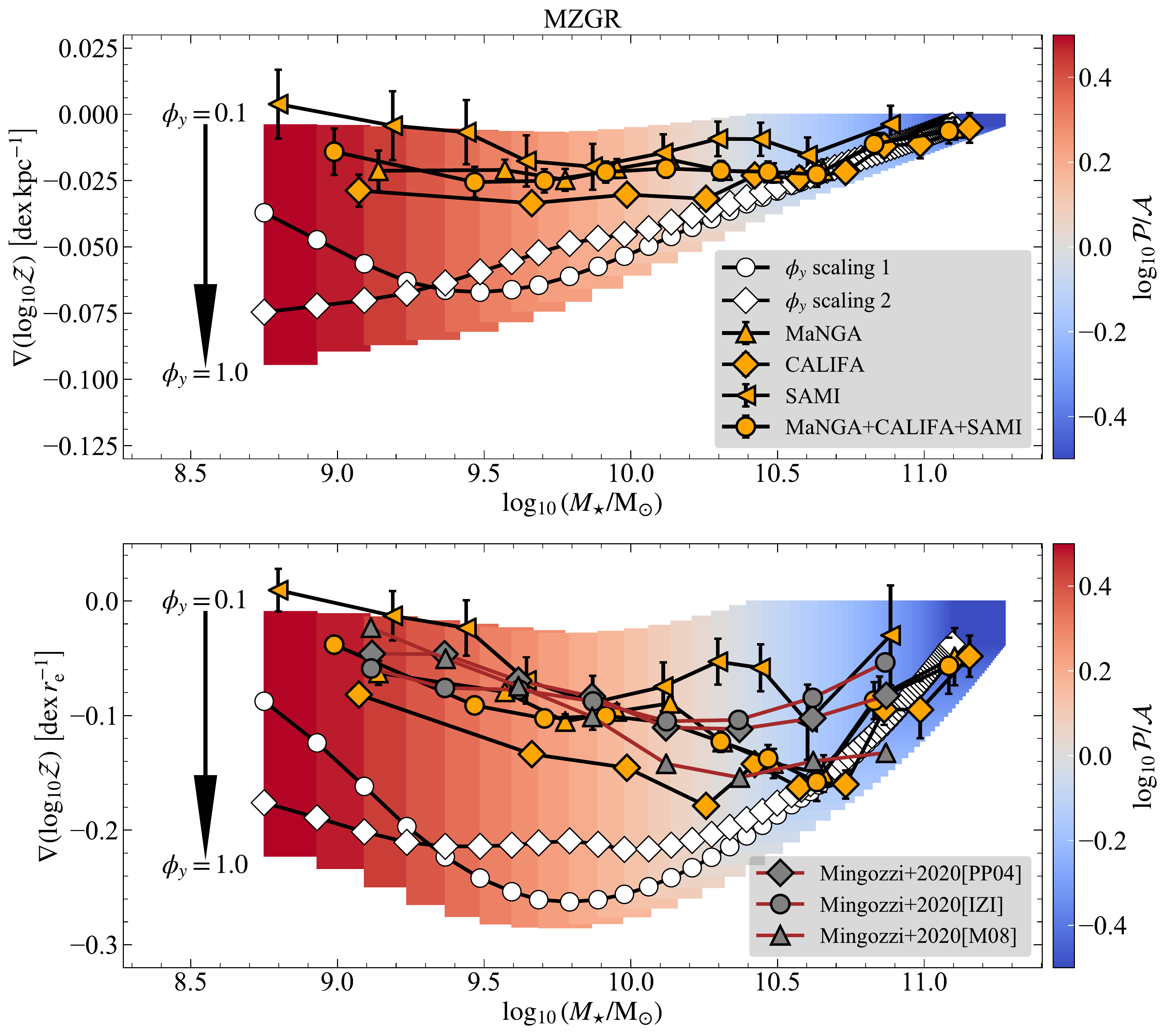}
\caption{The mass--metallicity gradient relation (MZGR) for the local Universe. The coloured band shows model predictions for different yield reduction factors, $\phi_y$ (note the opposite direction of the arrow as compared to \autoref{fig:mzr}), color-coded by the ratio of the Péclet number ($\mathcal{P}$) to cosmic accretion over diffusion ($\mathcal{A}$) in galaxies. The data to which we compare this model (orange points) are taken from a homogeneous analysis of metallicity gradients from the SAMI \protect\citep{2020aMNRAS.xxx..xxxP}, MaNGA \protect\citep{2017MNRAS.469..151B} and CALIFA \protect\citep{2014A&A...563A..49S} surveys, corrected for spatial resolution by \protect\cite{2020MNRAS.xxx..xxxA}. To give a sense of the systematic uncertainty, grey markers denote gradients measured with different metallicity calibrations (\protect\citealt[PP04]{2004MNRAS.348L..59P}, \protect\citealt[M08]{2008A&A...488..463M}, and \protect\citealt[IZI]{2015ApJ...798...99B}) for the MaNGA survey by \protect\cite{2020A&A...636A..42M}. Finally, we show model predictions with two possible empirical scalings of $\phi_y$ with $M_{\star}$ (white markers); these scalings are the same as in \autoref{fig:mzr}. The important conclusion from this plot is that metallicity gradients in local galaxies transition from the advection-dominated regime ($\mathcal{P} > \mathcal{A}$) to the accretion-dominated regime ($\mathcal{P} < \mathcal{A}$) as the stellar mass increases, and it is this transition that drives the shape of the MZGR. Note that the range in stellar mass covered by this figure is different than that shown in \autoref{fig:mzr}, due to differences in the mass ranges covered by the available observations.}
\label{fig:mzgr_mstar_PA}
\end{figure*}

We use the same metallicity distributions described in \autoref{s:MZR} to compute metallicity gradients. To be consistent with the procedure most commonly used in analysing observations, we obtain the gradient by performing a linear fit to $\log_{10}\,\mathcal{Z}$ from $0.5-2.5\,r_{\rm{e}}$ \citep[e.g.][]{2012A&A...538A...8S,2014A&A...563A..49S,2016A&A...587A..70S,2018MNRAS.479.5235P}.\footnote{To be consistent with observations, we only utilize metallicities till $2.5\,r_{\mathrm{e}}$ to measure the gradients, as opposed to $3\,r_{\mathrm{e}}$ that we use to measure $\overline{\mathcal{Z}}$.} Following the discussion on inverted gradients in \citet[Section~5.2.3]{2020aMNRAS.xxx..xxxS} and the uncertainty around them being in equilibrium, we restrict the model to produce only flat or negative gradients for the purposes of studying the MZGR. \autoref{fig:mzgr_mstar_PA} shows the MZGR from our model, again color-coded by the ratio of advection to accretion ($\mathcal{P}/\mathcal{A}$). The top and the bottom panels show the metallicity gradients in $\rm{dex\,kpc^{-1}}$ and $\mathrm{dex}\,r^{-1}_{\rm{e}}$ units, respectively. The spread, as for the MZR, is a result of $\phi_y$. The transition from the advection-dominated to the accretion-dominated regime, as in the MZR, is also visible in the MZGR. When the gradients are measured in $\mathrm{dex\,kpc^{-1}}$, this transition corresponds to the slight curvature in the MZGR that appears around $M_{\star} \sim 10^{10} - 10^{10.5}\,\rm{M_{\odot}}$ (top panel in \autoref{fig:mzgr_mstar_PA}). When they are measured in $\mathrm{dex}\,r^{-1}_{\mathrm{e}}$, it corresponds to the somewhat sharper curvature around the same stellar mass (bottom panel in \autoref{fig:mzgr_mstar_PA}). This finding is strong evidence for the links between the MZR and the MZGR, and also reveals that it is the same underlying physical mechanism that controls the shape of both. 

While the stellar mass of the accretion-advection transition influences the location at which our model curves bend, it is not the only factor that does so. The precise location of the bend is also sensitive to parameters like $\mathcal{Z}_{\rm{CGM}}$ and $\phi_y$, and both of the MZGR bend and the mass where $\mathcal{P}/\mathcal{A}=1$ depend weakly on the limits in $M_{\star}$ we select for smoothly interpolating between the dwarf and spiral regimes: for example, if we lower the threshold for spirals from $10^{10.5}\,\rm{M_{\odot}}$ to $10^{10}\,\rm{M_{\odot}}$, both shift to lower stellar mass. Similarly, if we increase the threshold for dwarfs from $10^{9}\,\rm{M_{\odot}}$ to $10^{9.5}\,\rm{M_{\odot}}$, both shift to higher stellar mass. However, irrespective of the interpolation limits in $M_{\star}$, both the curvature of the MZGR and the transition from $\mathcal{P} > \mathcal{A}$ to $\mathcal{P} < \mathcal{A}$ are always present. The existence of these features is a robust prediction of the model independent of uncertain parameter choices.

The physical origin for the behaviour of the MZGR is also the same as for the MZR: gradients are at their steepest when both of the processes for smoothing them -- accretion, $\mathcal{A}$, and inward advection of gas, $\mathcal{P}$, are at their weakest compared to metal production, $\mathcal{S}$. Diffusion also helps smooth gradients, but is always subdominant compared to either accretion or advection, as evidenced by the fact that we never have $\mathcal{P}<1$ and $\mathcal{A} < 1$ simultaneously. The point where advection and accretion are weakest is roughly where galaxies are transitioning from being advection-dominated, $\mathcal{P}>\mathcal{A}$, to accretion-dominated, $\mathcal{P}<\mathcal{A}$. We emphasise that, while the exact stellar mass at which this transition occurs can be somewhat sensitive to choices of model parameters (for example, the Toomre $Q$ of galactic discs), its existence is not; the bends in the coloured bands in \autoref{fig:mzgr_mstar_PA} that describe our model always occur irrespective of our parameter choices. Additionally, note that the minimum of the model MZGR is not always coincident with $\mathcal{P}/\mathcal{A}=1$; the position of the minimum is dependent on the model parameters, in particular, $\phi_y$.

In \autoref{fig:mzgr_mstar_PA} we also plot MZGRs from the MaNGA \citep{2017MNRAS.469..151B}, CALIFA \citep{2014A&A...563A..49S,2016A&A...587A..70S} and SAMI \citep{2018MNRAS.479.5235P,2020aMNRAS.xxx..xxxP} surveys, homogenized and corrected for spatial resolution by \cite{2020MNRAS.xxx..xxxA}. We adopt the $\rm{dex\,kpc^{-1}}$ values from \cite{2020MNRAS.xxx..xxxA}, and convert to $\mathrm{dex}\,r^{-1}_{\rm{e}}$ following the $r_{\mathrm{e}}$--$M_{\star}$ scaling relations from \cite{2014ApJ...788...28V} to be consistent with our assumptions elsewhere\footnote{The qualitative trend of the MZGR remains the same for the $\mathrm{dex}\,r^{-1}_{\rm{e}}$ gradients reported by \cite{2020MNRAS.xxx..xxxA} as compared to the ones shown in the bottom panel of \autoref{fig:mzgr_mstar_PA} using the scaling relation between $r_{\mathrm{e}}$--$M_{\star}$, with a change in the overall normalisation of the metallicity. We have also verified that the $r_{\rm{e}}$ we find from \cite{2014ApJ...788...28V} is in very good agreement with that measured in, for example, the SAMI sample we use.}. We also overplot results from MaNGA based on three different metallicity calibrations by \cite{2020A&A...636A..42M}: \citet[PP04]{2004MNRAS.348L..59P}, \citet[M08]{2008A&A...488..463M}, and \citet[IZI]{2015ApJ...798...99B}. 

The first thing to notice is that the qualitative trend found in the data is in good agreement with that predicted by our model: gradients are steepest at $M_\star \sim 10^{10} - 10^{10.5}$ M$_\odot$, and flatten at both lower and higher masses. However, the location of the curvature in the data and the model differ by as much as $0.5-1\,\rm{dex}$ in stellar mass. This is not surprising given the uncertainties in the parameters that affect the curvature, as discussed above (e.g., interpolation limits in $M_{\star}$, our constant adopted value of $\mathcal{Z}_{\rm{CGM}}$, and the scaling of $\phi_y$ with $M_{\star}$). Moreover, it is important to recall that the data themselves are not fully secure, due to uncertainties caused by the choice of metallicity diagnostic; \citet[their Figure~11]{2020aMNRAS.xxx..xxxP} show that the exact mass at which the MZGR bends depends on which diagnostic is used to determine the metallicity, and that these variations are reduced but still persist even after the diagnostics are homogenised. Thus, it is presently difficult to accurately determine the location of the curvature, especially given its mildness. Nonetheless, the presence of a bend seems to be robust in the data, as it is in our model.

Second, we see that similar to the MZR, this comparison of the model to the observed MZGR reveals that low-mass galaxies prefer low $\phi_y$. However, the spread due to $\phi_y$ in the MZGR at the high-mass end is quite narrow; thus, gradients in massive galaxies are not particularly sensitive to $\phi_y$, although the data suggests higher $\phi_y$ for the MZGR in massive galaxies (note the inverted arrows for $\phi_y$ on \autoref{fig:mzgr_mstar_PA} as compared to \autoref{fig:mzr}). Our findings on $\phi_y$ being ineffective at setting gradients in massive galaxies is consistent with earlier works \citep[e.g.,][]{2013MNRAS.434.1531F}. However, our proposed explanation for the flattening of gradients in massive galaxies based on the advection-to-accretion transition differs from these studies that attributed the observed flattening to saturation of ISM metallicities \citep{1991MNRAS.251...84P,2017MNRAS.468..305M}, radially-varying star formation efficiency \citep{2019MNRAS.487..456B}, or past mergers \citep{2010ApJ...710L.156R,2011MNRAS.417..580P,2013MNRAS.434.1531F}.

In \autoref{fig:mzgr_mstar_PA}, we also plot model predictions using the two scalings of $\phi_y$ with $M_{\star}$ that we described in \autoref{s:MZR}. These scalings are able to reproduce the high mass end of the MZGR, and yield a qualitative trend similar to that seen in the data, but quantitatively the predicted gradients from the scalings are steeper than that observed at the low mass end. In retrospect, this is not entirely unexpected given the uncertainties in the two approaches, and the fact that these scalings are sensitive to the absolute metallicity (see \aref{s:app_phiy_mstar}). Judging from \autoref{fig:mzgr_mstar_PA}, we slightly prefer scaling 2, since it is closer to the observations at intermediate stellar masses; we revisit the comparison between the two scalings in \autoref{s:MZR-MZGR}. Nevertheless, the fact that both the MZR and the MZGR suggest a qualitatively similar scaling between $\phi_y$ and $M_{\star}$ is an encouraging sign of consistency. However, it is difficult to derive quantitative similarities given the uncertainties in these empirical scalings.

\subsection{Comparison with previous work}
\label{s:compare_mzgr}
Only a handful of models exist in the literature that focus on gas phase metallicity gradients rather than global metallicities \citep{2013MNRAS.435.2918M, 2013ApJ...765...48J, 2015MNRAS.448.2030H,2015MNRAS.451..210C, 2015MNRAS.450..342K,2016MNRAS.455.2308P,2017MNRAS.467.1154S,2021arXiv210106833K}, and even fewer that actually study the local MZGR or its equivalent \citep{2018MNRAS.476.3883L,2019MNRAS.489.1436L,2019MNRAS.487..456B}. Of these, the models by \cite{2018MNRAS.476.3883L} and \cite{2019MNRAS.487..456B} are closest in spirit to ours.\footnote{\citet{2019MNRAS.489.1436L} focus only on low-mass satellites, so our results are not easily comparable.} Quantitative comparison between our results and those of \citeauthor{2018MNRAS.476.3883L} is challenging, because they do not quote measurements in $\rm{dex/kpc}$ or equivalent. Examining their plots, it seems that they also find slightly steeper gradients for intermediate mass galaxies, consistent with our findings. Similarly, \citeauthor{2019MNRAS.487..456B} find that observed gradients in local dwarfs and spirals are best reproduced by a model where the star formation timescale at each radius is proportional to the local orbital period. For massive galaxies, this scaling is quite similar to that in the \citet{2018MNRAS.477.2716K} galaxy model that is embedded in our metallicity model, and thus at first glance is also consistent with our findings. However, there remain substantial differences between our model and those of \citeauthor{2018MNRAS.476.3883L} and \citeauthor{2019MNRAS.487..456B}. Neither of these studies include the effects of radial inflow or metal diffusion. Neither adopt our approach of systematically varying the highly-uncertain yield reduction factor $\phi_y$: \citeauthor{2018MNRAS.476.3883L} adopt a parameterised, time-dependent functional form that they tune in order to match stellar and gas metallicity gradient data, while \citeauthor{2019MNRAS.487..456B} assume that the ISM and outflow metallicities are equal ($\phi_y =1$ in our terminology), contrary to our findings and inconsistent with the available observational evidence \citep{2002ApJ...574..663M,2016MNRAS.457.2642S,2018MNRAS.481.1690C,2019ApJ...877..120T,2020MNRAS.499..193K}. Finally, both sets of authors explicitly fit their model parameters to the data, whereas we do not except while introducing the second scaling in $\phi_y$. Thus, it is unclear to what extent the agreement between the models is simply a matter of their being enough adjustable parameters to make them behave similarly.

In addition to analytic models, semi-analytic models like L-Galaxies 2020 have also investigated the local MZGR, finding somewhat flatter gradients for massive galaxies as compared to low mass galaxies \citep{2020arXiv201104670Y}. The authors attribute their findings to inside-out star formation that increases the gas phase metallicity in the inner disc in massive galaxies. In the outer disc in these galaxies, \citeauthor{2020arXiv201104670Y} either find metal-rich accretion from the CGM that enhances the metallicity (their `modified' model), or metal-poor accretion that dilutes the metallicity at every radius (their `default' model). The combined effect is to produce flatter metallicity profiles in massive galaxies in each case. They further conclude that flattening of the metallicity profiles in massive local galaxies is expected regardless of the mass-loading factors of outflows. Thus, their explanations for the trends seen in the local MZGR are consistent with the findings of our model. It is worth noting that while working with an earlier version of L-Galaxies, \cite{2013MNRAS.434.1531F} found relative metal enrichment of outflows to be more important than advection in driving gas phase metallicities. These authors also find a trend in the MZGR consistent with \citeauthor{2020arXiv201104670Y} and ours.

It is also helpful to compare our results to simulations that have studied the local MZGR. For example, both \cite{2016MNRAS.456.2982T} and \cite{2017MNRAS.466.4780M} find slightly flatter gradients for massive local galaxies in their simulations, consistent with our model and available observations. The EAGLE simulations \citep{2015MNRAS.446..521S} find that metallicity gradients in their simulated galaxies are systematically shallower at $z=0$ than those observed in the local Universe due to high star formation efficiency at all radii \citep[Figure~11]{2019MNRAS.482.2208T}. As a result, the MZGR predicted from their simulations does not show any clear trends with the stellar mass. On the other hand, the local MZGR produced by the IllustrisTNG50 simulations \citep{2019MNRAS.490.3196P,2019MNRAS.490.3234N} is in very good quantitative agreement with that produced by our model, both in terms of the mean gradient and the scatter in gradients at a given stellar mass \citep[Figure~8]{2020arXiv200710993H}. These authors suspect that gradients flatten in massive local galaxies due to AGN feedback and increasing galaxy size. While the latter of the two is consistent with the findings of \cite{2020aMNRAS.xxx..xxxS}, the primary driver of flatter gradients in massive galaxies in our model is due to the increasing role of metallicity dilution by cosmic accretion.

\begin{figure*}
\includegraphics[width=\columnwidth]{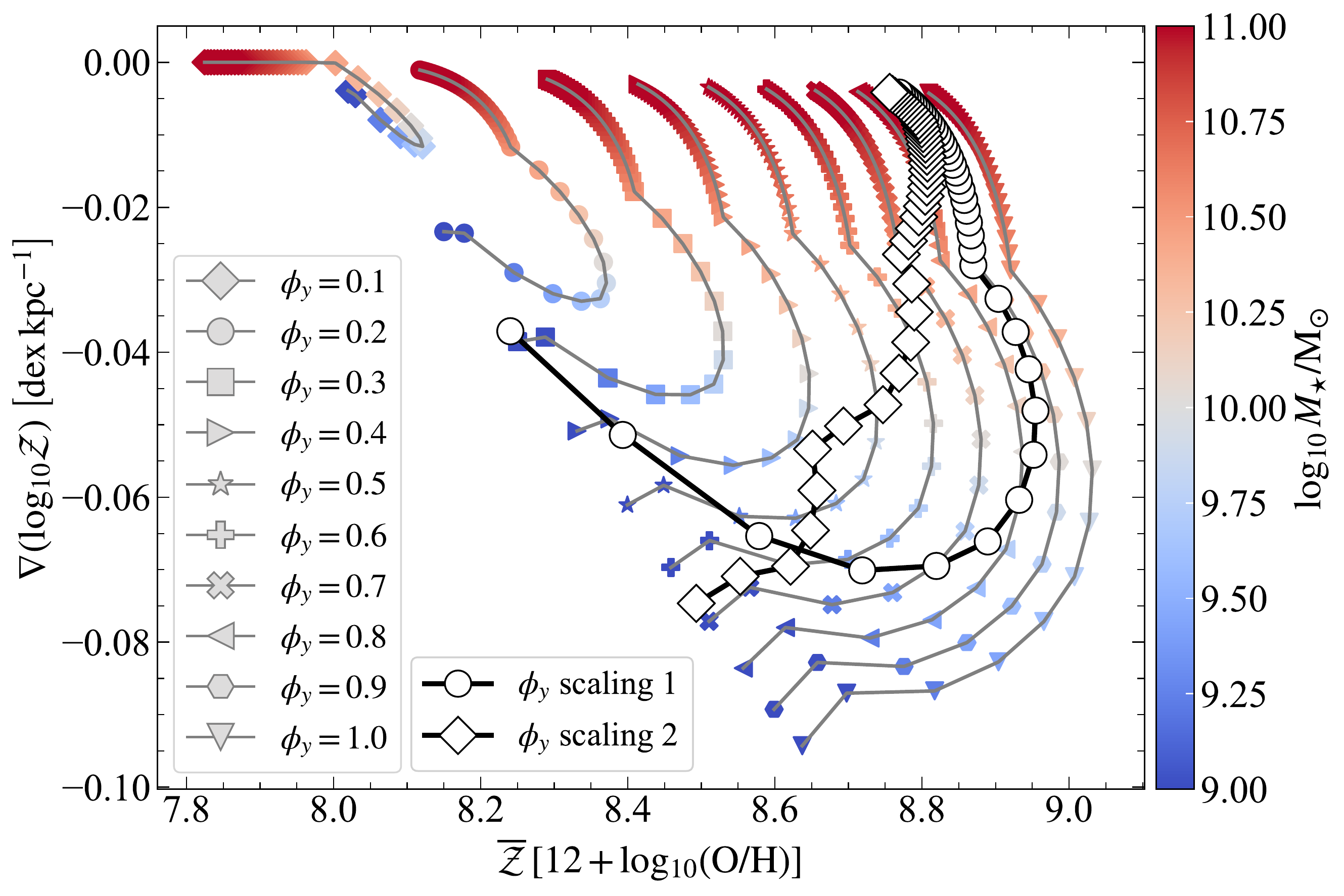}
\includegraphics[width=\columnwidth]{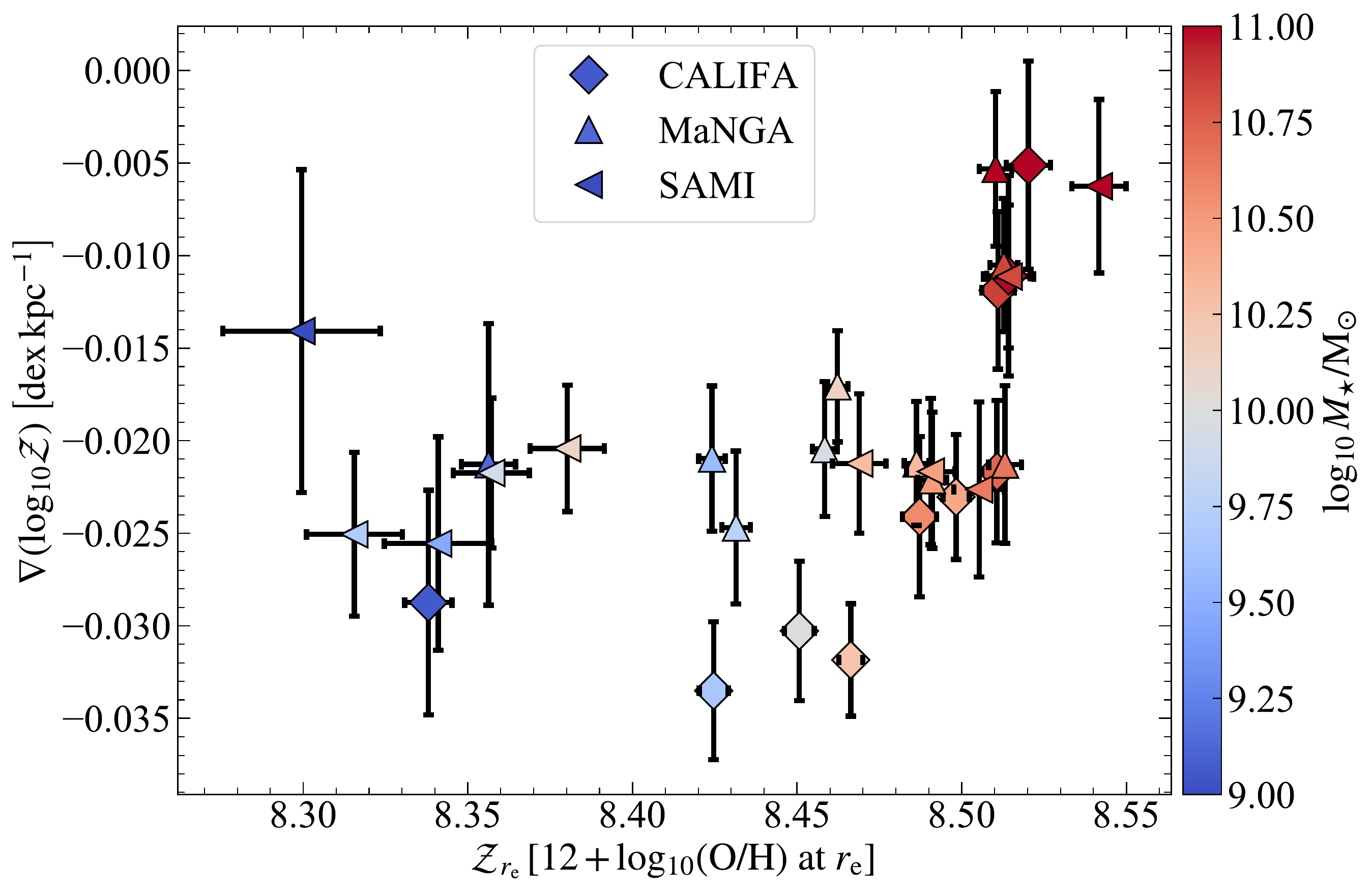}
\caption{\textit{Left panel:} MZGR--MZR space from the model for the local Universe, defined by the metallicity gradient (in $\mathrm{dex\,kpc^{-1}}$) as a function of the global (SFR-weighted) galaxy metallicity (defined as in \autoref{eq:Zmean}). Points are color-coded by stellar mass, and different curves represent the different yield reduction factor, $\phi_y$, which describes the metal-enrichment of galactic outflows. Both the MZR and the MZGR predict a scaling of $\phi_y$ with $M_{\star}$ such that low-mass galaxies prefer low $\phi_y$, implying that these galaxies lose a higher proportion of the metals they produce to winds, as compared to massive galaxies. Also overlaid are the two empirical scalings of $\phi_y$ with $M_{\star}$ that are shown in \autoref{fig:mzr} and \autoref{fig:mzgr_mstar_PA}. The bend seen at intermediate masses corresponds to the advection-to-accretion transition identified in \autoref{fig:mzr} and \autoref{fig:mzgr_mstar_PA}. The range in $M_{\star}$ covered in this plot is slightly different from that in \autoref{fig:mzr} and \autoref{fig:mzgr_mstar_PA}. \textit{Right panel:} Mean metallicity gradients as a function of metallicity at the effective radius $r_{\rm{e}}$ in the CALIFA, MaNGA and SAMI surveys that we adopt from \protect\cite{2020MNRAS.xxx..xxxA}. The observations show a similar bend compared to the predictions of the model in the MZR--MZGR space. Note, however, the differences in the axes ranges between this panel and the left panel, reflecting the difficulty of putting metallicity measurements at specific radius ($r_{\rm{e}}$) and ``global'' metallicities on a common scale. The trends in the model as well as the data in the MZR-MZGR space remain qualitatively similar when the gradients are plotted in units of $\mathrm{dex}\,r^{-1}_{\mathrm{e}}$ instead of $\mathrm{dex}\,\mathrm{kpc^{-1}}$.}
\label{fig:mzr-mzgr}
\end{figure*}


\section{The MZR--MZGR relation}
\label{s:MZR-MZGR}
In this section, we introduce a new way of looking at galaxy metallicities, by studying the MZR$-$MZGR correlation space. The two-fold motivation behind this is to: (1.) understand how global metallicities correlate with metallicity gradients in galaxies, because this can inform us about the correlations between global and internal dynamics of galaxies, and (2.) given that both the MZR and the MZGR require similar scaling of $\phi_y$ with $M_{\star}$ to reproduce the observations, we can study the relative importance of $\phi_y$ for both of these relations. An additional advantage of studying this parameter space is that it can be constructed both in observations and simulations. 

In order to construct the MZR--MZGR correlation space in the model, we simply plot $\nabla (\log_{10}\mathcal{Z})$ from \autoref{fig:mzgr_mstar_PA} as a function of $\overline{\mathcal{Z}}$ from \autoref{fig:mzr}. We show this in the left panel of \autoref{fig:mzr-mzgr}, where we color-code the model points by $M_{\star}$, with different curves corresponding to different $\phi_y$. Note that the range in $M_{\star}$ is slightly different in this plot as compared to that in \autoref{fig:mzr} and \autoref{fig:mzgr_mstar_PA}; thus, there are some differences visible in this plot as compared to previous figures. It is clear from this plot that $\phi_y$ has two distinct effects. At the high-mass end, it simply shifts the overall metallicity $-$ $\overline{\mathcal{Z}} \propto \phi_y$ $-$ without significantly affecting the gradient. At the low-mass end, it affects the overall metallicity, but also affects the gradient, by making it steeper for larger $\phi_y$. It is also clear that the relationship between $\overline{\mathcal{Z}}$ and $\nabla (\log_{10}\mathcal{Z})$ is non-monotonic because of the same $\mathcal{P}/\mathcal{A}$ split we have seen in the MZR and the MZGR, \textit{i.e.,} there are two typical branches where $\overline{\mathcal{Z}}$ and $\nabla (\log_{10}\mathcal{Z})$ change monotonically with respect to one another, but the curves bend when galaxies transition from the advection-dominated to the accretion-dominated regime. Irrespective of the value of $\phi_y$, this bend always occurs around $10^{10-10.5}\,\rm{M_{\odot}}$ because it is dictated by the ratio $\mathcal{P}/\mathcal{A}$ crossing unity. To demonstrate the robustness of this feature, we also overplot results for the two empirical scalings of $\phi_y$ with $M_{\star}$ that we discussed in previous sections. We see that both empirical scalings also produce a bend in the $\overline{\mathcal{Z}}-\nabla(\log_{10}\mathcal{Z})$ plane, but with rather different amounts of curvature. Thus, a generic prediction of our model is that galaxies should lie along a bent track in $\overline{\mathcal{Z}}-\nabla(\log_{10}\mathcal{Z})$ space, with one arm closer to vertical and one closer to horizontal, but we cannot predict the exact shape of this track without a better understanding of how $\phi_y$ varies with $M_\star$. The trends in the model we identify in the MZR-MZGR space remain qualitatively the same when the gradients are plotted in units of $\mathrm{dex}\,r^{-1}_{\mathrm{e}}$, so we do not discuss them separately.

We create a parameter space similar to that above by plotting the measured metallicity gradients as a function of the measured gas phase metallicity at $r_{\rm{e}}$ from \cite{2020MNRAS.xxx..xxxA}\footnote{The conversion from metallicity at $r_{\rm{e}}$ to mean metallicity is non-trivial and suffers considerable calibration uncertainties, both in the observations and in the model (which does not use $r_e$ as a parameter or make an independent prediction of its location in the disc), which is why we do not attempt to create an MZR from the same observations for which we have the MZGR to directly study the MZR--MZGR space.}. We show this in the right panel of \autoref{fig:mzr-mzgr}, color-coded with $M_{\star}$. The main takeaway from this figure is that the data shows a qualitatively similar bend at $M_{\star} \sim 10^{10.5}\,\rm{M_{\odot}}$ as the model. While this is not a one-to-one comparison between the model and the data given the former uses global metallicity whereas the latter uses metallicity at a specific location in the disc, we expect the qualitative trend (\textit{i.e.,} the presence of the bend) to be robust given the findings in the previous sections. Similar to our observations in \autoref{s:MZGR}, we find that scaling 2 better reproduces the trend seen in the data. Further, like the model, the same trends in the data are also present when the gradients are plotted in units of $\mathrm{dex}\,r^{-1}_{\mathrm{e}}$. Thus, the model is able to identify and recover the presence of this bend in the metallicity--metallicity gradient space, and sets clear predictions for future work that will enable us to re-construct this space and facilitate a direct comparison with the model.

Hence, in addition to our findings in \autoref{s:MZR} and \autoref{s:MZGR}, we conclude that metal-enriched outflows play a crucial role in setting both the MZR and the MZGR for low-mass galaxies, while for high-mass galaxies, outflows play a significant role only for the MZR.

\section{Conclusions}
\label{s:conclusions}
In this work, we present a physical explanation for the observed relation between the stellar mass and the gas phase metallicity gradient (MZGR) for galaxies in the local Universe, using the recently-developed first-principles model of gas phase metallicity gradients in galaxies given by \cite{2020aMNRAS.xxx..xxxS}. We show that the shape of the MZGR is driven by the balance between metal advection and production for low-mass galaxies, and between cosmic accretion and metal production for massive galaxies. The point where the MZGR begins to curve as the galaxy mass increases corresponds to the transition of galaxies from the advection-dominated to the accretion-dominated regime. Additionally, the best match between the model and the data naturally recovers the expected dependence of the MZGR on metal-enrichment of galactic outflows: low-mass galaxies have more metal-rich winds as compared to massive galaxies, implying that metals in low mass galaxies are not well-mixed with the ISM before ejection. This is in good agreement with observations \citep{2002ApJ...574..663M,2018MNRAS.481.1690C} and simulations \citep{2018ApJ...869...94E,2018ApJ...867..142C, 2020ApJ...899..108T}.

We also present the first joint explanation for the mass-metallicity relation (MZR) and the MZGR. We find that in addition to the model successfully reproducing both the MZR and the MZGR, it has two primary commonalities: (1.) the curvature observed in both the MZR and the MZGR around a stellar mass $M_\star \approx 10^{10-10.5}\,\rm{M_{\odot}}$ have the same underlying cause, which is the shift between radial advection (in low-mass galaxies) and cosmological accretion (in more massive galaxies) as the dominant agent supplying metal-poor gas to galaxy centres, and (2.) both the MZR and the MZGR produced by the model predict that supernova-produced metals in low-mass galaxies are largely ejected before mixing with the ISM, while metals in high-mass galaxies are well-mixed with the ISM. The fact that the MZR and MZGR results are qualitatively consistent with each other is evidence for the links between global and spatially-resolved galaxy properties, though our ability to check this quantitatively is currently limited by the large uncertainties in observed metallicites. 

In studying these relations, we also introduce a new way of characterizing gas phase metallicities via the MZR--MZGR correlation space. We find that the relation between the global metallicity and metallicity gradient in galaxies is non-monotonic, and bends as a result of the advection-to-accretion transition identified above. We also retrieve this bend in the available data (in metallicity gradient--metallicity at $r_{\rm{e}}$ space), although limitations due to the mismatch between model and data techniques prevent us from constructing the observed MZR--MZGR space exactly as we do for the model. Moreover, the MZR--MZGR space also disentangles the relative importance of metal-enriched outflows for the global metallicities and metallicity gradients: while metal-enrichment of the outflows significantly influences both the global metallicity and metallicity gradients in low-mass galaxies, in massive galaxies only the absolute metallicity is sensitive to the properties of the outflows, and gradients are flat regardless of outflow metallicity.

\section*{Acknowledgements}
We thank the anonymous reviewer for their feedback, which helped to improve the paper. We also thank Lisa Kewley for going through a preprint of this paper and providing comments, and Roland Crocker for useful discussions. PS is supported by the Australian Government Research Training Program (RTP) Scholarship. MRK and CF acknowledge funding provided by the Australian Research Council (ARC) through Discovery Projects DP190101258 (MRK) and DP170100603 (CF) and Future Fellowships FT180100375 (MRK) and FT180100495 (CF). MRK is also the recipient of an Alexander von Humboldt award. PS, EW and AA acknowledge the support of the ARC Centre of Excellence for All Sky Astrophysics in 3 Dimensions (ASTRO 3D), through project number CE170100013. CF further acknowledges an Australia-Germany Joint Research Cooperation Scheme grant (UA-DAAD). JCF is supported by the Flatiron Institute through the Simons Foundation. Analysis was performed using \texttt{numpy} \citep{oliphant2006guide,2020arXiv200610256H} and \texttt{scipy} \citep{2020NatMe..17..261V}; plots were created using \texttt{Matplotlib} \citep{Hunter:2007}. This research has made extensive use of NASA's Astrophysics Data System (ADS) Bibliographic Services. The ADS is a digital library portal for researchers in astronomy and physics, operated by the Smithsonian Astrophysical Observatory (SAO) under a NASA grant. This research has also made extensive use of \texttt{Wolfram|Alpha} and \texttt{Mathematica} for numerical analyses, and the image-to-data tool \texttt{WebPlotDigitizer}. 

\section*{Data Availability Statement}
No data were generated for this work.



\bibliographystyle{mnras}
\bibliography{references}

\appendix

\section{Scaling of the yield reduction factor with stellar mass}
\label{s:app_phiy_mstar}
Comparing our analytic model with observations of the MZR and the MZGR discussed in the main text suggests a scaling of the yield reduction factor $\phi_y$ with stellar mass $M_{\star}$. In this appendix, we explore ways to directly retrieve this scaling using two different methods. The two scalings introduced below capture the qualitative essence of how $\phi_y$ should scale with $M_{\star}$, albeit with significant uncertainties. 

\begin{figure}
\includegraphics[width=\columnwidth]{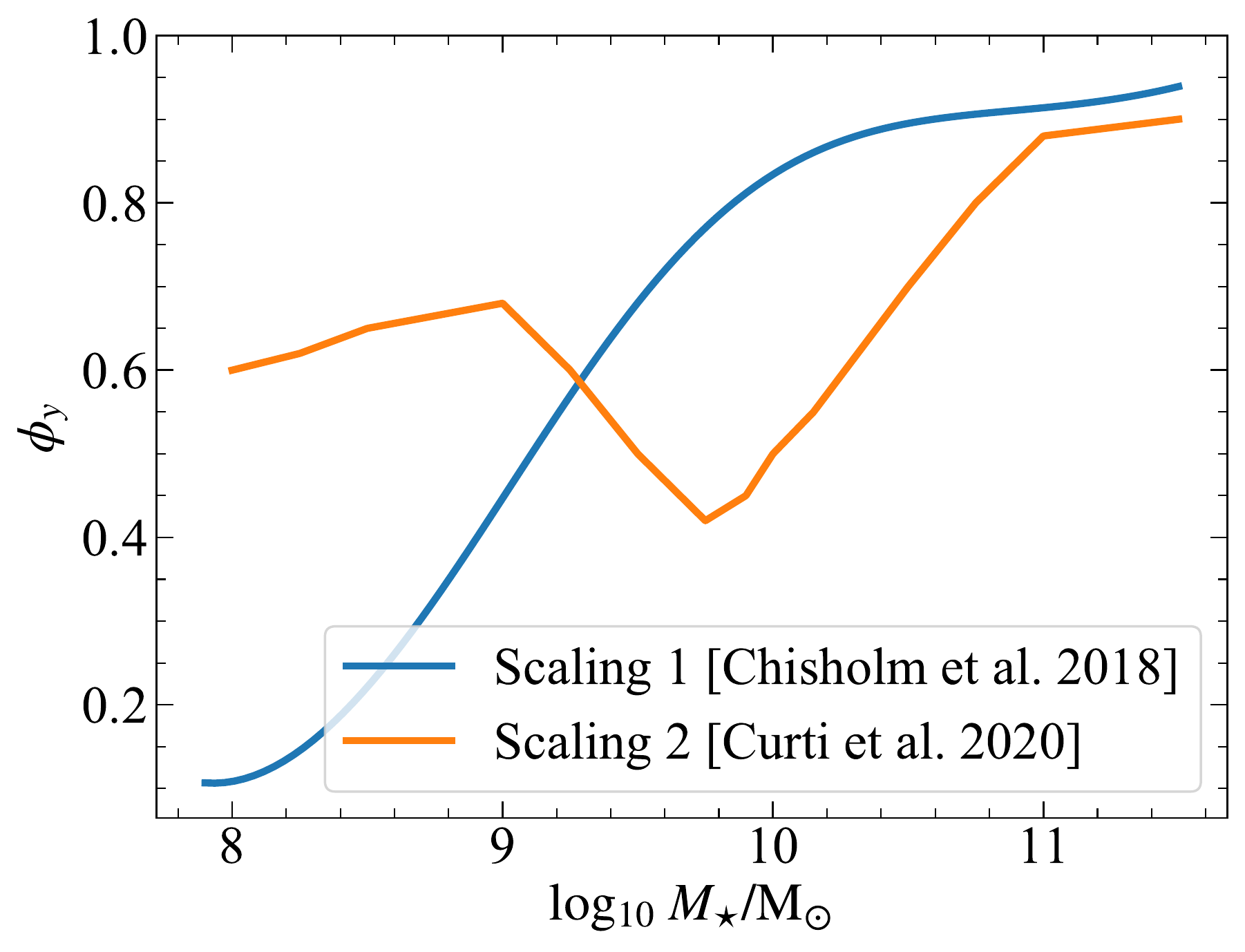}
\caption{Scalings of the yield reduction factor $\phi_y$ with $M_{\star}$, obtained using the two approaches described in \aref{s:app_phiy_mstar}. Scaling 1 is from observations \protect\citep{2018MNRAS.481.1690C} whereas scaling 2 is from the best match between the model MZR and the \protect\cite{2020MNRAS.491..944C} MZR.}
\label{fig:app_phiy_mstar}
\end{figure}

\begin{enumerate}
    \item \textit{Scaling 1:} We make use of the observations reported in \cite{2018MNRAS.481.1690C} to derive a scaling of $\phi_y$ with $M_{\star}$. The authors report on the ratio of the wind to the ISM metallicity, $\mathcal{Z}_w/\mathcal{Z}$, as well as the metal mass loading factor for galaxies of different masses. We perform a linear fit to their data to obtain $\mathcal{Z}_w/\mathcal{Z}$ as a function of $M_{\star}$. To find how the metal mass loading factor varies as a function of $M_{\star}$, we use the scaling provided by \cite{2002MNRAS.330...69D} which provides the best fit to the data. Then, we use $\mathcal{Z}_w/\mathcal{Z}$ and the metal mass loading factor to find the mass loading factor $\mu$ as a function of $M_{\star}$. Using $\mathcal{Z}_w/\mathcal{Z}$ and $\mu$, it is straightforward to compute $\phi_y$ \citep[equations 10 and 13]{2020aMNRAS.xxx..xxxS}
    \begin{equation}
        \phi_y = 1 - \mu\mathcal{Z} \frac{\rm{Z_{\odot}}}{y}\left(\frac{\mathcal{Z}_w}{\mathcal{Z}}-1\right)\,,
    \label{eq:chisholm}
    \end{equation}
    where $y$ is the yield of metals from core collapse supernovae. Before we proceed further, it is important to point out the caveats of this approach. Firstly, \citeauthor{2018MNRAS.481.1690C} only observed 7~galaxies across a wide range of $M_{\star}$ ($\sim 10^7-10^{11}\,\rm{M_{\odot}}$), so the coverage in stellar mass is very sparse. Secondly, the ISM metallicities for the galaxies used in \citeauthor{2018MNRAS.481.1690C} are non-homogeneous; for example, some are stellar metallicities and some are gas metallicities. Thirdly, some galaxies in this dataset are undergoing mergers, and show diluted metallicities as compared to isolated galaxies of the same mass. Keeping these caveats in mind, and noting that $\phi_y$ is sensitive to the absolute value of $\mathcal{Z}$ as we see from \autoref{eq:chisholm}, we simply increase the ISM metallicities quoted in \citeauthor{2018MNRAS.481.1690C} by 0.3~dex, which has the effect of bringing them into closer alignment with the observed MZR; without this increment, the least and the most massive galaxies in the sample ($M_{\star} \approx 10^7$ and $10^{10.7}\,\rm{M_{\odot}}$, respectively) would have a metallicity $\mathcal{Z}=0.03$ and $0.5$, respectively, placing them well below the observed MZR. We do not re-scale the ratio $\mathcal{Z}_w/\mathcal{Z}$ because it is not sensitive to the absolute value of $\mathcal{Z}$. With this adjustment, we show the resulting scaling of $\phi_y$ with $M_\star$ in \autoref{fig:app_phiy_mstar}. This is our first model scaling.
    
    \item \textit{Scaling 2:} In this approach, we simply find the best match between the model MZRs and the \cite{2020MNRAS.491..944C} MZR by eye, where we take the latter to be the representative MZR in the local Universe. We note that there is no particular reason to prefer the latter MZR over other available MZRs, especially given the uncertainties in the absolute normalisation of metallicities. However, for the sake of developing a scaling of $\phi_y$ with $M_{\star}$ from this approach, we will continue with this MZR. We plot the resulting scaling in \autoref{fig:app_phiy_mstar}. Interestingly, while the general trend of $\phi_y$ increasing with $M_{\star}$ still holds, we find an inflection at intermediate masses where $\phi_y$ is the lowest. However, we do not place great weight on this finding, given the large uncertainties in both the choice of MZR and its absolute value. From the standpoint of our model predictions, the main difference between this scaling and our first scaling is that this scaling gives a shallower trend in $\phi_y$ with $M_\star$, such that $\phi_y$ reaches a minimum value of only $\approx 0.5$ even for very low-mass galaxies.
\end{enumerate}

\bsp	
\label{lastpage}
\end{document}